\documentclass[iop]{emulateapj_mod}

\def\snr{SNR~CTA~1}
\def\ver{VER~J0006+729}
\def\rxj{RX~J0007.0+7303}
\def\psr{PSR~J0007+7303}
\def\EG{3EG~J0010+7309}
\def\centroid{\ensuremath{00^h\  06^m\ 26^s,\ +72^\circ\ 59^\prime\  01.0^{\prime\prime}}}

\def\etal{et~al.}

\def\rosat{{\it ROSAT}}
\def\egret{{\it EGRET}}
\def\asca{{\it ASCA}}
\newcommand\chandra{{\it Chandra}}

\newcommand\xmm{{\it XMM-Newton}}

\newcommand\veritas{{\it VERITAS}}
\newcommand\VERITAS{{\it VERITAS}}

\newcommand\fermi{{\it Fermi}}

\shortauthors{Aliu et al.}

\begin{document}

\submitted{To appear in The Astrophysical Journal}
\title{Discovery of TeV Gamma-ray Emission from CTA 1 by VERITAS}

\author{
E.~Aliu\altaffilmark{1},
S.~Archambault\altaffilmark{2},
T.~Arlen\altaffilmark{3},
T.~Aune\altaffilmark{4},
M.~Beilicke\altaffilmark{5},
W.~Benbow\altaffilmark{6},
A.~Bouvier\altaffilmark{4},
J.~H.~Buckley\altaffilmark{5},
V.~Bugaev\altaffilmark{5},
A.~Cesarini\altaffilmark{7},
L.~Ciupik\altaffilmark{8},
E.~Collins-Hughes\altaffilmark{9},
M.~P.~Connolly\altaffilmark{7},
W.~Cui\altaffilmark{10},
R.~Dickherber\altaffilmark{5},
C.~Duke\altaffilmark{11},
J.~Dumm\altaffilmark{12},
V.~V.~Dwarkadas\altaffilmark{13},
M.~Errando\altaffilmark{1},
A.~Falcone\altaffilmark{14},
S.~Federici\altaffilmark{15,16},
Q.~Feng\altaffilmark{10},
J.~P.~Finley\altaffilmark{10},
G.~Finnegan\altaffilmark{17},
L.~Fortson\altaffilmark{12},
A.~Furniss\altaffilmark{4},
N.~Galante\altaffilmark{6},
D.~Gall\altaffilmark{18},
G.~H.~Gillanders\altaffilmark{7},
S.~Godambe\altaffilmark{17},
E.~V.~Gotthelf\altaffilmark{19},
S.~Griffin\altaffilmark{2},
J.~Grube\altaffilmark{8},
G.~Gyuk\altaffilmark{8},
D.~Hanna\altaffilmark{2},
J.~Holder\altaffilmark{20},
G.~Hughes\altaffilmark{15},
T.~B.~Humensky\altaffilmark{21},
P.~Kaaret\altaffilmark{18},
O.~Kargaltsev\altaffilmark{22,23},
N.~Karlsson\altaffilmark{12},
Y.~Khassen\altaffilmark{9},
D.~Kieda\altaffilmark{17},
H.~Krawczynski\altaffilmark{5},
F.~Krennrich\altaffilmark{24},
M.~J.~Lang\altaffilmark{7},
K.~Lee\altaffilmark{5},
A.~S~Madhavan\altaffilmark{24},
G.~Maier\altaffilmark{15},
P.~Majumdar\altaffilmark{3},
S.~McArthur\altaffilmark{25,*},
A.~McCann\altaffilmark{26},
P.~Moriarty\altaffilmark{27},
R.~Mukherjee\altaffilmark{1,*},
T.~Nelson\altaffilmark{12},
A.~O'Faol\'{a}in de Bhr\'{o}ithe\altaffilmark{9},
R.~A.~Ong\altaffilmark{3},
M.~Orr\altaffilmark{24},
A.~N.~Otte\altaffilmark{28},
N.~Park\altaffilmark{25},
J.~S.~Perkins\altaffilmark{29,30},
M.~Pohl\altaffilmark{16,15},
H.~Prokoph\altaffilmark{15},
J.~Quinn\altaffilmark{9},
K.~Ragan\altaffilmark{2},
L.~C.~Reyes\altaffilmark{31},
P.~T.~Reynolds\altaffilmark{32},
E.~Roache\altaffilmark{6},
M.~Roberts\altaffilmark{33},
D.~B.~Saxon\altaffilmark{20},
M.~Schroedter\altaffilmark{6},
G.~H.~Sembroski\altaffilmark{10},
P.~Slane\altaffilmark{34},
A.~W.~Smith\altaffilmark{17},
D.~Staszak\altaffilmark{2},
I.~Telezhinsky\altaffilmark{16,15},
G.~Te\v{s}i\'{c}\altaffilmark{2},
M.~Theiling\altaffilmark{10},
S.~Thibadeau\altaffilmark{5},
K.~Tsurusaki\altaffilmark{18},
J.~Tyler\altaffilmark{2},
A.~Varlotta\altaffilmark{10},
V.~V.~Vassiliev\altaffilmark{3},
S.~Vincent\altaffilmark{15},
M.~Vivier\altaffilmark{20},
S.~P.~Wakely\altaffilmark{25},
T.~C.~Weekes\altaffilmark{6},
A.~Weinstein\altaffilmark{24},
R.~Welsing\altaffilmark{15},
D.~A.~Williams\altaffilmark{4},
B.~Zitzer\altaffilmark{35}
}

\altaffiltext{*}{Corresponding authors: \\S.~McArthur: smcarthur@ulysses.uchicago.edu, \\R.~Mukherjee: muk@astro.columbia.edu}
\altaffiltext{1}{Department of Physics and Astronomy, Barnard College, Columbia University, NY 10027, USA}
\altaffiltext{2}{Physics Department, McGill University, Montreal, QC H3A 2T8, Canada}
\altaffiltext{3}{Department of Physics and Astronomy, University of California, Los Angeles, CA 90095, USA}
\altaffiltext{4}{Santa Cruz Institute for Particle Physics and Department of Physics, University of California, Santa Cruz, CA 95064, USA}
\altaffiltext{5}{Department of Physics, Washington University, St. Louis, MO 63130, USA}
\altaffiltext{6}{Fred Lawrence Whipple Observatory, Harvard-Smithsonian Center for Astrophysics, Amado, AZ 85645, USA}
\altaffiltext{7}{School of Physics, National University of Ireland Galway, University Road, Galway, Ireland}
\altaffiltext{8}{Astronomy Department, Adler Planetarium and Astronomy Museum, Chicago, IL 60605, USA}
\altaffiltext{9}{School of Physics, University College Dublin, Belfield, Dublin 4, Ireland}
\altaffiltext{10}{Department of Physics, Purdue University, West Lafayette, IN 47907, USA }
\altaffiltext{11}{Department of Physics, Grinnell College, Grinnell, IA 50112-1690, USA}
\altaffiltext{12}{School of Physics and Astronomy, University of Minnesota, Minneapolis, MN 55455, USA}
\altaffiltext{13}{Department of Astronomy and Astrophysics, University of Chicago, Chicago, IL, 60637}
\altaffiltext{14}{Department of Astronomy and Astrophysics, 525 Davey Lab, Pennsylvania State University, University Park, PA 16802, USA}
\altaffiltext{15}{DESY, Platanenallee 6, 15738 Zeuthen, Germany}
\altaffiltext{16}{Institute of Physics and Astronomy, University of Potsdam, 14476 Potsdam-Golm, Germany}
\altaffiltext{17}{Department of Physics and Astronomy, University of Utah, Salt Lake City, UT 84112, USA}
\altaffiltext{18}{Department of Physics and Astronomy, University of Iowa, Van Allen Hall, Iowa City, IA 52242, USA}
\altaffiltext{19}{Columbia Astrophysics Laboratory, Columbia University, New York, NY 10027, USA}
\altaffiltext{20}{Department of Physics and Astronomy and the Bartol Research Institute, University of Delaware, Newark, DE 19716, USA}
\altaffiltext{21}{Physics Department, Columbia University, New York, NY 10027, USA}
\altaffiltext{22}{Department of Astronomy, University of Florida, 205 Bryant Space Center, Gainesville, FL 32611, USA}
\altaffiltext{23}{Department of Physics, The George Washington University, Washington, DC 20052, USA}
\altaffiltext{24}{Department of Physics and Astronomy, Iowa State University, Ames, IA 50011, USA}
\altaffiltext{25}{Enrico Fermi Institute, University of Chicago, Chicago, IL 60637, USA}
\altaffiltext{26}{Kavli Institute for Cosmological Physics, University of Chicago, Chicago, IL 60637, USA}
\altaffiltext{27}{Department of Life and Physical Sciences, Galway-Mayo Institute of Technology, Dublin Road, Galway, Ireland}
\altaffiltext{28}{School of Physics and Center for Relativistic Astrophysics, Georgia Institute of Technology, 837 State Street NW, Atlanta, GA 30332-0430}
\altaffiltext{29}{CRESST and Astroparticle Physics Laboratory NASA/GSFC, Greenbelt, MD 20771, USA.}
\altaffiltext{30}{University of Maryland, Baltimore County, 1000 Hilltop Circle, Baltimore, MD 21250, USA.}
\altaffiltext{31}{Physics Department, California Polytechnic State University, San Luis Obispo, CA 94307, USA}
\altaffiltext{32}{Department of Applied Physics and Instrumentation, Cork Institute of Technology, Bishopstown, Cork, Ireland}
\altaffiltext{33}{Eureka Scientific, Inc., Oakland, CA 94602, USA}
\altaffiltext{34}{Harvard-Smithsonian Center for Astrophysics, 60 Garden Street, Cambridge, MA 02138, USA}
\altaffiltext{35}{Argonne National Laboratory, 9700 S. Cass Avenue, Argonne, IL 60439, USA}

\begin{abstract}

We report the discovery of TeV gamma-ray emission coincident with the shell-type radio supernova remnant (SNR) CTA~1 using the \VERITAS\ gamma-ray observatory. 
The source, \ver, was detected as a 6.5 standard deviation excess over background and shows
 an extended morphology, approximated by a two-dimensional Gaussian of semi-major (semi-minor) axis 
 $0^\circ.30$ ($0^\circ.24$) 
and a centroid $5'$ from the Fermi gamma-ray pulsar \psr\ and its X-ray pulsar wind nebula (PWN). The photon spectrum is well described by a power-law $dN/dE = N_0(E/3\ \rm TeV)^{-\Gamma}$, with a differential spectral index of $\Gamma= 2.2\pm 0.2_{\rm stat} \pm 0.3_{\rm sys}$, and normalization $N_0= (9.1\pm 1.3_{\rm stat} \pm 1.7_{\rm sys})\times 10^{-14}$ cm$^{-2}$ s$^{-1}$ TeV$^{-1}$.  The integral flux, $F_\gamma = 4.0\times10^{-12} \mathrm{\ erg\ cm^{-2}\ s^{-1}}$ above 1~TeV, corresponds to 0.2\% of the pulsar spin-down power at 1.4~kpc. 
 The energetics, co-location with the SNR, and the relatively small extent of the  TeV emission strongly argue for the PWN origin of the TeV photons. 
We consider the origin of the TeV emission in CTA 1.

\end{abstract}



\section{Introduction}\label{sec:intro}

There are many possible associations of gamma-ray sources with supernova remnants (SNRs).  These gamma rays could come from shock acceleration in the shell, a pulsar associated with the SNR, or a pulsar wind nebula (PWN) surrounding the pulsar.  
Such gamma-ray/SNR associations date back to {\it COS~B} observations of SNRs coincident with OB stellar associations \citep{Montmerle1979}. 
Observations of the Galaxy by \egret\ in the energy range 30~MeV--30~GeV
revealed $\sim19$ unidentified sources at low Galactic latitudes that were found to be spatially correlated  with mostly shell-type SNRs \citep{Torres2003}. 
One such source was \EG, with a relatively small 95\% error circle of $28^{\prime}$ \citep{Hartman1999}, that was spatially coincident  with the \snr\ (G119.5+10.2) and the X-ray point source \rxj, which was postulated to be a pulsar \citep{Brazier1998}. 
The association between \EG\ and \rxj\  was found to be plausible, given the lack of flux variability seen in \EG, its hard spectral index ($\Gamma = 1.58\pm 0.18$ between 70~MeV and 2~GeV), and its similarity with other known pulsars detected by \egret\  \citep{Brazier1998}. 

CTA~1 is a composite SNR, discovered by \citet{HarrisPASP1960}, with a shell-type structure in the radio band and 
a center-filled morphology at X-ray energies. The radio shell is incomplete towards the north-west (NW) of the remnant, possibly due to rapid expansion of the shock into a lower-density region 
 \citep{Pineault1993}. 
The distance to \snr\ as derived from the associated HI shell is $d=1.4\pm 0.3$ kpc \citep{Pineault1997}, the SNR 
 age is estimated to be $\sim 1.3\times 10^4$~yr \citep{Slane2004}, and the diameter of its radio shell is 
 $\sim 1^\circ.8$ \citep{Sieber1981}. 


Archival X-ray observations of \snr\ 
in the 5--10~keV band show non-thermal diffuse emission of low surface brightness in the center of the remnant, likely corresponding to a PWN 
driven by an active neutron star \citep{Slane1997}. 
The neutron star candidate \rxj\ 
is a faint source located at the brightest part of the synchrotron emission 
\citep{Seward1995}. A \chandra\ image of \rxj\ provides further evidence that this source is an energetic rotation-powered pulsar, resolving a central point source, a compact nebula, and a bent jet \citep{Halpern2004}. An initial 
observation with \xmm\ in 2002 
found the X-ray spectrum of the central source to be consistent with that of a neutron star, although no pulsations were detected \citep{Slane2004}. 
Based on these initial X-ray observations, the 
spin-down luminosity of the underlying 
pulsar was 
 estimated to be in the range $10^{36}-10^{37}$ ergs~s$^{-1}$, supporting the identification of the \egret\ source \EG\ as a pulsar \citep{Slane2004,Halpern2004}. 

Eventually, 
a search for pulsed GeV emission from CTA~1 
using the data from the \fermi\  {\sl Gamma-Ray Space Telescope} revealed a highly significant 316~ms signal, confirming the origin of \EG\ \citep{Abdo2008CTA1}. The spin-down power was determined to be $\sim 4.5\times 10^{35}$~erg~s$^{-1}$, 
 which is sufficient to power the pulsar wind nebula 
 \citep{Slane2004}. Following the \fermi\ discovery of the gamma-ray pulsar, a deep 130~ks observation of 
  \rxj\ was carried out with \xmm\  to characterize the timing behavior. 
  The X-ray 
signal of \psr\
 was discovered at a statistical significance of $4.7\sigma$ in the 0.5-2~keV band, out of phase with the gamma-ray pulse 
  \citep{Caraveo2010}. Similar to 
  Geminga \citep{Halpern1992} and PSR~J1836+5925 
  \citep{Halpern2007, Abdo2010Geminga}, \psr\ is also radio quiet. 
 GeV emission in the off-pulse phase interval has also recently been detected 
  by \fermi\ \citep{Abdo2012CTA1}.


Many galactic TeV sources appear to be 
 associated with pulsars via their wind nebulae, 
comprised  of relativistic wind particles confined by the pressure of the surrounding medium \citep{Gaensler2006}.  The initially highly supersonic wind  terminates in a shock which can be associated with axisymmetric, toroidal  structures often seen in the X-ray images of 
PWNe \citep[e.g.,][]{Kargaltsev2008}.
PWNe now represent the most populous class of Galactic TeV emitters \citep{Hinton2010}. 
  The non-thermal emission seen in PWNe from the radio up to gamma rays 
 below a few GeV or less is generally interpreted as synchrotron radiation from the accelerated leptons. The emission observed at higher energies, up to several TeV, can be produced via inverse Compton (IC) scattering of these same high-energy electrons with background photons  (e.g. the cosmic microwave background (CMB), infrared radiation from dust, starlight, and synchrotron photons) \citep{Atoyan1996}. 
 Alternatively, hadronic mechanisms may also be responsible for the TeV emission, in which case the wind should be composed 
 of relativistic hadrons 
 that collide with the ambient medium 
and produce pions, with the TeV emission coming from $\pi^0$ decay. 
 To date, however, there has been no solid evidence requiring a large contribution to the gamma-ray emission from such hadronic processes. 
 
 PWNe experience several stages of evolution \citep[e.g.][]{Gaensler2006}. At an early stage the pulsar wind freely expands into the SN ejecta. For a slowly moving pulsar the PWN is approximately centered on the pulsar while for a supersonically moving pulsar the PWN will take a cometary shape. At later times, 
  the PWN is compressed by the reverse SNR shock and may be displaced significantly from the pulsar if the reverse shock is asymmetric. Such crushed and displaced PWNe have been dubbed relic PWNe. 

The X-ray and gamma-ray observations of CTA~1 suggest that the extended non-thermal emission around the gamma-ray pulsar is a synchrotron PWN. 
 Motivated by these observations, model calculations by \citet{Zhang2009} 
suggested that 
 the TeV 
 emission is largely produced 
 by the PWN, 
  and that the level of emission should 
  be detectable at TeV energies by \veritas.  For 
  IC  scattering off the PWN electrons, 
   \citet{Zhang2009} predicted a gamma-ray flux $F_\gamma (1-30 {\rm\ TeV}) \sim 1.1\times 10^{-12}$ erg cm$^{-2}$ s$^{-1}$.  
Previous TeV observations of CTA~1 by the earlier 
imaging atmospheric Cherenkov telescopes gave 
 upper limits, as follows:  $2.64\times 10^{-11}$ photons cm$^{-2}$ s$^{-1}$ ($E>250$~GeV) by CAT \citep{Khelifi2001}, 
$1.25 \times 10^{-11}$ photons cm$^{-2}$ s$^{-1}$ ($E>620$~GeV) by Whipple \citep{Hall2001}, 
 and $1.09\times 10^{-12}$ photons cm$^{-2}$ s$^{-1}$ ($E>1.3$~TeV) by HEGRA  \citep{Rowell2003}.

In this paper, we report the \VERITAS\ detection of 
 TeV  emission from the central region of  CTA~1.

\vspace{0.1cm}

\section{VERITAS Instrument \& Observations}\label{sec:observations}

The Very Energetic Radiation Imaging Telescope Array System (\veritas) uses ground-based detection techniques pioneered by its predecessor, 
 the Whipple 10m Telescope 
 \citep{Weekes1989}, to explore the Universe in very high-energy (VHE) gamma rays from $\sim 100$~GeV to $\sim 30$~TeV. 
The \veritas\ telescope array 
consists of four 12-m diameter Davies-Cotton telescopes and is located at the basecamp of the Fred Lawrence Whipple Observatory (FLWO) in southern Arizona \citep{Holder2011}. 
%
Flashes of Cherenkov light from gamma-ray and cosmic-ray showers are focused by a set of mirrors onto a camera located in the focal plane of each telescope.  
 Each 
 camera comprises 499 photomultiplier tube pixels and light concentrators arranged in a hexagonal pattern with a total field-of-view 
  of 3$^\circ$.5. 
  Stereoscopic imaging of showers from multiple viewing angles allows the determination of the shower core location relative to the array using simple geometric reconstruction techniques which rely on the fact that the major axes of the shower images are projections of the shower axis.  
The combined instrument 
 has an angular resolution of $<0^\circ.1$ 
  (68\% containment) and energy resolution of $15-20\%$ for energies $>200$~GeV. It can detect a source with a flux of 1\% of the steady Crab Nebula VHE flux at a 5 standard deviation significance level in less than 30~hours \citep{Ong2009}.

A three-level trigger system is used to help eliminate background noise.  The first trigger occurs at the pixel level, requiring the signal to reach a 50~mV threshold (corresponding to 4--5 photoelectrons) set by a constant fraction discriminator (CFD).  The second,  telescope-level trigger requires at least three adjacent pixels passing the CFD trigger to form an image. A third, array-level trigger requires simultaneous Cherenkov images in at least two telescopes, within a 50~ns 
time window, which then causes 
 a readout of the 500~MSample/s Flash-ADC data acquisition system for each pixel.

CTA~1 was observed over two epochs. The first set of observations spanned from September 2010 to January 2011, with a total livetime of 25~hours 
39~min after data-quality selection based on weather conditions and hardware status. An additional 15~hours 
36~min of quality-selected data were taken from September to December 2011. 
 Observations were taken in ``wobble'' mode \citep{Fomin1994}, in which the telescope pointing is offset from the source position by some angular distance. This method allows for simultaneous collection of data and estimation of the background. 
 Due to the extended nature of the remnant and expected extension of the PWN, an offset distance of $0^\circ.7$ 
  was used, larger than the typical \VERITAS\ distance of $0^\circ.5$. 
To decrease bias, the offset direction was varied between each 20~minute run while maintaining the same offset distance, alternating between the north, south, east and west directions (in the equatorial coordinate system). 
 Observations were taken in a narrow range of zenith angles, $40-47^\circ$, with an average of $42^\circ.5$ 
  for the full dataset.  
All of the data presented here were taken with all four telescopes in the array. 

\section{Analysis}\label{sec:analysis}

The CTA~1 data were processed using standard \veritas\ analysis techniques, as described in \citet{Acciari2008}. The cosmic-ray background was suppressed efficiently by parametrizing the recorded shower images by their principal 
 moments \citep{Hillas1985}, and the shower direction and impact parameter were reconstructed from these images, using stereoscopic methods \citep[see e.g.][]{Aharonian1997, Krawczynski2006}. 
Gamma-ray/hadronic shower separation is achieved through selection criteria (\textit{cuts}). Based upon the predicted spectrum of \citet{Zhang2009},  two sets of standard cuts were used. These cuts were 
 optimized on simulations for sources of $\sim5\%$   of the Crab Nebula flux and with moderate and hard spectral indices ($\sim$ -2.5 and -2.0, respectively). For these cuts, 
at least three of the telescopes in the array had to have images recorded in the camera, with at least 1200 digital counts ($\sim 240$ photo-electrons) for the hard-cut analysis and 500 digital counts ($\sim 95$ photo-electrons) for the moderate-cut analysis. 
Cuts were also applied to the mean scaled length (\textit{MSL}), mean scaled width (\textit{MSW}), and integrated charge in the signal (\textit{size}). 
 Finally a cut was applied on $\theta$, the angular distance in the field of view from the reconstructed arrival direction of the event to the putative source location.  A cut of $\theta<0^\circ.09$ ($\theta<0^\circ.23$) 
 was used for a point-source (extended-source) search, with the size of the extended-source cut selected \textit{a priori}. 
For the analysis presented here, the cuts for the moderate- and hard-spectra analysis are $MSW < 0.35$ and $MSL < 0.7$.

The background was estimated using the ring background model \citep[e.g., see][]{Berge2007},
 with a ring of mean radius $0^\circ.7$ 
and a background to source area ratio of 8.0. 
Regions in the field of view containing stars of $B$ magnitude brighter than 6.0 were excluded from the background estimation in order to reduce systematic errors. 
The statistical significance of the excess was calculated using Equation~17 of \citet{LiMa1983}.
The energy threshold for this analysis after applying the moderate (hard) cuts is $\sim550$~GeV (1~TeV) at a zenith angle of $45^\circ$, with a systematic error  
of about  20\% on the energy estimation.  Two independent analysis packages, as described by \citet{Cogan2008} and \citet{Daniel2008}, were used to reproduce the results presented here on CTA~1.  

\section{Results}\label{sec:results}

\begin{figure}[tbp]
\begin{center}
\includegraphics[angle=270,width=\columnwidth]{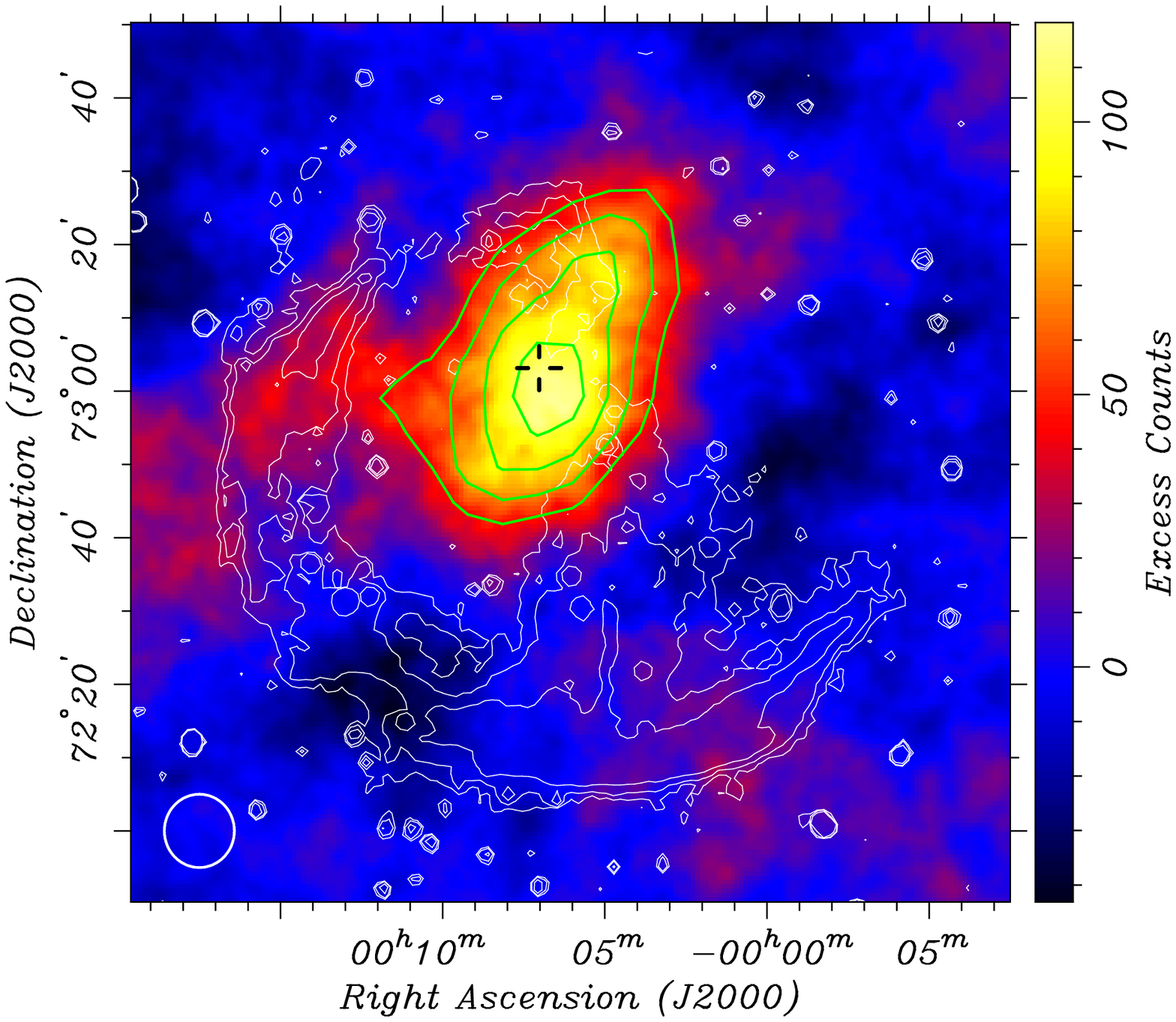}
\end{center}
\vspace{-0.3cm}
\caption{\veritas\ excess map of the region around \snr\ using a hard-spectrum analysis. 
The color scale indicates excess gamma-ray events within an integration radius of 0$^\circ$.23. 
The circle at the lower left corner shows the size of the point-spread function (68\% containment). 
The radio contours at 1420 MHz \citep{Pineault1997} are overlaid in white, 
 showing the SNR shell. The \veritas\ significance contours at 3, 4, 5, and $6\sigma$ are shown in green. 
The cross marks the position of the pulsar \citep{Abdo2008CTA1}, located $5'\pm 2'$ from the centroid of TeV emission. North is up and east is to the left.}
\label{fig:hardextended_radioexcess} 
\vspace{-0.2cm}
\end{figure}

Figure~\ref{fig:hardextended_radioexcess} shows the TeV excess map of the region of the sky around CTA~1. 
The hard-spectrum, extended-source analysis produced an 
excess with a significance of 7.5 standard deviation ($\sigma$)  pre-trials, in a 
 search region of radius $0^\circ.4$ 
 around the pulsar \psr, within the radio shell of the SNR CTA~1. Accounting for the two sets of applied cuts with two different integration radii, and determining the {\sl a priori} trials factor by tiling the search region 
  with 0$^\circ$.04 
  bins~\citep{Aharonian2006}, we estimate the post-trials significance of detection to be $6.5\sigma$. 
  Overlaid on the TeV 
  image are the high-resolution radio contours at 1420~MHz, obtained using the Dominion Radio Astrophysical Observatory (DRAO) Synthesis Telescope, and the Effelsberg 100-m telescope, 
showing bright radio arcs visible to the south and east, with an incomplete shell in the northwest, possibly due to the breakout of the SNR blast wave into a medium of lower density \citep{Pineault1993,Pineault1997}.

\begin{figure}[tbp]
\begin{center}
\includegraphics[width=1.1\columnwidth]{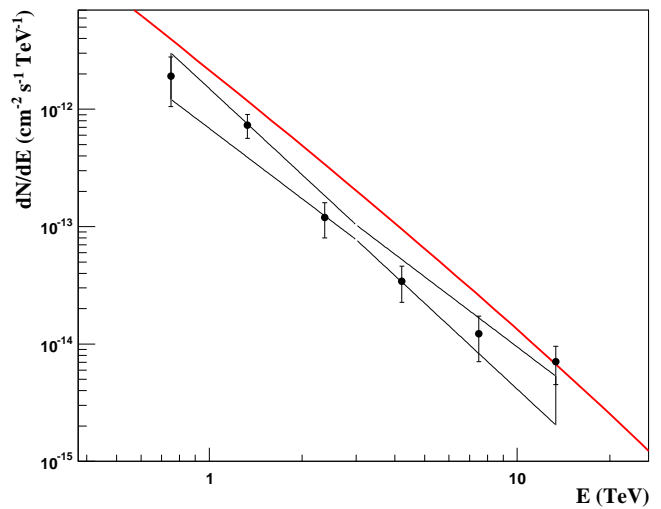}
\end{center}
\vspace{-0.4cm}
\caption{\veritas\ differential gamma-ray spectrum of CTA~1.  The black butterfly shows the uncertainties of the best-fit power-law model. 
 The red line marks the flux predicted by \citet[Figure 4]{Zhang2009}. The errors are statistical only.}
\label{fig:veritas_sed} 
\end{figure}

For spectral analysis, the moderate-spectrum cuts were used in order to provide the lowest energy threshold for the analysis. 
The differential photon spectrum 
above 500~GeV is shown in Figure~\ref{fig:veritas_sed}, with spectral data points listed in Table~\ref{tab:veritas_spectralPoints}.  The spectrum is generated with the reflected-region 
 background model \citep{Berge2007} with 41~hours 15~min 
 of quality-selected data.  The spectrum can be fit with a power-law of the form $dN/dE =  N_0 (E/3\mathrm{\ TeV})^{-\Gamma}$, with 
$\Gamma = 2.2 \pm 0.2_{\mathrm{stat} }\pm 0.3_{\mathrm{sys}}$
and 
$N_0 = (9.1 \pm 1.3_{\mathrm{stat}} \pm 1.7_{\mathrm{sys}})\times 10^{-14}
\mathrm{\ cm^{-2}\ s^{-1}\ TeV^{-1}}$.  The integral 
energy
flux 
above 1 TeV,
$F_\gamma = 4.0\times10^{-12} \mathrm{\ erg\ cm^{-2}\ s^{-1}}$,
corresponds to 0.2\% of the pulsar spin-down luminosity at 1.4~kpc and $\sim 4\%$ 
of the steady TeV gamma-ray emission from the Crab Nebula. 

\begin{table}[tbp]
\caption{Differential flux measurements of CTA~1 with \VERITAS. The errors are statistical only.}
\vspace{-0.3cm}
\begin{center}
\begin{tabular}{ccc}
\hline
Energy Range & Flux & Significance \\
~[TeV] & [cm$^{-2}$ s$^{-1}$ TeV$^{-1}$] & [$\sigma$] \\ \hline
0.56 -- 1.00 & $(1.9 \pm 0.9) \times 10^{-12}$ & 2.3 \\
1.00 -- 1.78 & $(7.3 \pm 1.7) \times 10^{-13}$ & 4.5 \\
1.78 -- 3.16 & $(1.2\pm 0.4) \times 10^{-13}$ & 3.3 \\
3.16 -- 5.62 & $(3.4 \pm 1.2) \times 10^{-14}$ & 3.1 \\ 
5.62 -- 10.00 & $(1.2 \pm 0.5) \times 10^{-14}$ & 2.5 \\ 
10.00 -- 17.78 & $(7.1 \pm 2.5) \times 10^{-15}$ & 2.8 \\ \hline
\end{tabular}
\end{center}
\label{tab:veritas_spectralPoints} 
\vspace{-0.2cm}
\end{table}

\pagebreak

\subsection{Morphology}
Figure~\ref{fig:hardextended_radioexcess} shows that the extent of the TeV gamma-ray emission region in CTA~1 exceeds the 
 point-spread function (PSF; 68\% 
 containment radius of the events coming from a point source) of \veritas. 
 In order to estimate the extent of the source, an asymmetric two-dimensional 
 Gaussian is fit to the acceptance-corrected uncorrelated map of excess events binned in 0$^\circ$.05 
  bins.  Although the shape and extent of the emission is likely more complex than a simple asymmetric Gaussian, as a first approximation, it still provides a statistically reasonable estimate of the source extent. 
  Due to the finite resolution of the detector, the emission we see is a convolution of the real source and the PSF describing the system. Accounting for the PSF of the instrument, the resulting 1$\sigma$ angular extent is $0^\circ.30 \pm 0^\circ.03$ 
  along the semi-major axis and $0^\circ.24  \pm 0^\circ.03$ 
   along the semi-minor axis, with an orientation angle of 
$17^\circ.4 \pm 15^\circ.8$ 
west of north.  
We note that this is a sensitivity-limited measurement. 

The fitted centroid location is \centroid\ (J2000), which is 5 arcmin from \psr.  Therefore, the \VERITAS\ source name is \ver.  The statistical uncertainty in the centroid position is $0^\circ.09$ in RA and $0^\circ.04$ 
in declination. 
The systematic uncertainty in the position due to the
telescope pointing error is 
50''.
 
\subsection{Archival X-ray Analysis}
\begin{figure}[tbp]
\begin{center}
\includegraphics[angle=270,width=\columnwidth]{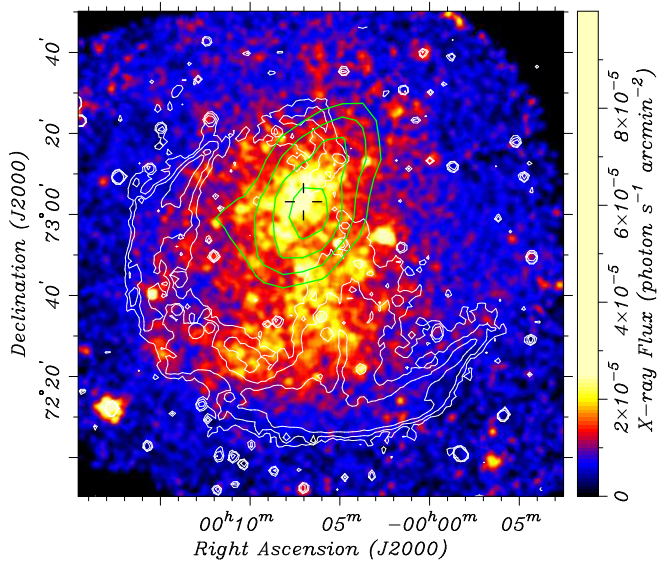} 
\end{center}
\vspace{-0.3cm}
\caption{\rosat\ X-ray image (0.5--2.0~keV) of the \snr\ shown in equatorial coordinates. The cross marks the location of the X-ray point source \rxj\ and the \fermi\ \psr. The SNR shell is shown by the 1420 MHz radio contours \citep{Pineault1997}, overlaid in white. The \veritas\ significance contours at 3, 4, 5, and $6\sigma$ are shown in green.}
\label{fig:rosat_xray} 
\end{figure}

\begin{figure}[tbp]
\begin{center}
\includegraphics[angle=270,width=\columnwidth]{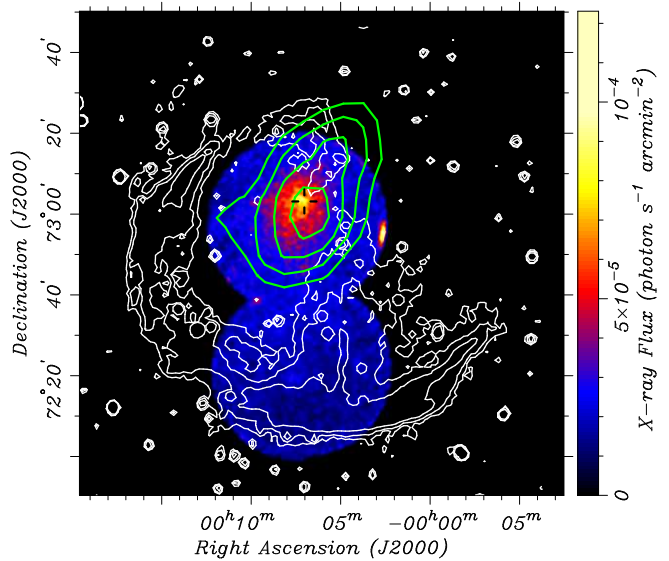} 
\end{center}
\vspace{-0.3cm}
\caption{\asca\ GIS image (4--10~keV) of the \snr, using the same field of view as Figures~\ref{fig:hardextended_radioexcess} and~\ref{fig:rosat_xray}. The position of \psr\ is marked by the cross. The 1420 MHz radio contours are shown in white. The \veritas\ significance contours at 3, 4, 5, and $6\sigma$ are shown in green. }
\label{fig:ASCA_xray} 
\end{figure}

Figure~\ref{fig:rosat_xray} shows the exposure-corrected, smoothed \rosat\ PSPC X-ray image of the region around CTA~1. 
 The 0.5--2~keV \rosat\ image reveals  
 a center-filled morphology, with a faint compact source located at the peak of the brightness distribution. The cross in the image marks 
 the location of the X-ray point source \rxj\ and the \fermi\ pulsar J0007+7303.  The pulsar is located close to the center of the extended TeV source with $\sim 5'$ 
 offset from the peak of the TeV surface brightness. 
Figure~\ref{fig:ASCA_xray} shows the non-thermal X-ray image from \asca\ in the 4--10~keV band \citep{Roberts2001}, along with the TeV contours. 
While the non-thermal emission and the TeV emission are both clearly extended with centroids separated by only 5 arcmin, the limited angular resolution and signal-to-noise ratio in the \VERITAS\ data do not allow for a more rigorous comparison.

\begin{figure*}[tbp]
\begin{center}
\includegraphics[scale=0.5,angle=0]{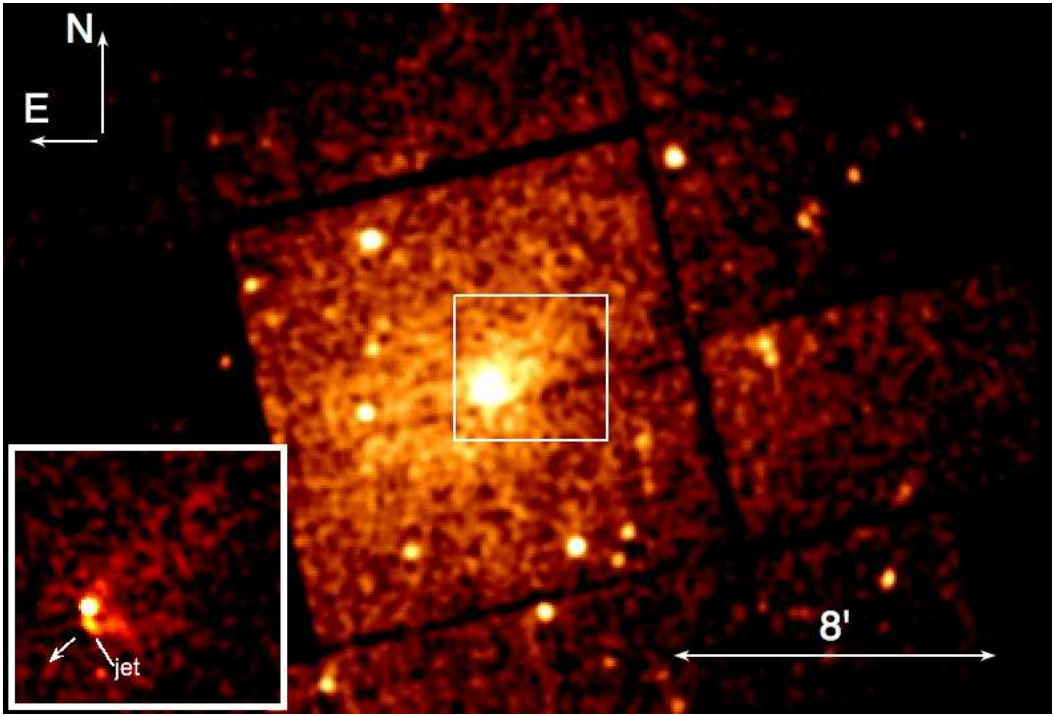} 
\end{center}
\caption{\xmm\ 93~ks EPIC/MOS1+2  image (0.5--10 keV; pixel size $4''$; smoothed with the $r=12''$ Gaussian 
kernel) of \psr\ 
 and its vicinity showing the X-ray PWN. The inset is the higher resolution 50~ks \chandra\ ACIS image (0.5--8 keV; pixel size $2''$, smoothed with the $r=4''$ Gaussian 
 kernel). The arrow in the inset image shows the likely direction of the pulsar proper motion (based on the shape of the compact PWN).}
\label{fig:xray_pwn} 
\end{figure*}

Figure~\ref{fig:xray_pwn} 
is the smoothed \xmm\ 
image of the vicinity of \psr\ (ObsID 0604940101, PI P.~Caraveo) showing the X-ray PWN. The inset shows the smoothed \chandra\ image (ObsID 3835, PI J.~Halpern)
 revealing a compact nebula and bent jet attached to the point source, 
 along with diffuse emission at larger scales. 
The \chandra\ jet is particularly apparent in the analysis presented by \citet{Halpern2004}, where it is estimated that the 
 \chandra\ point source accounts for $\sim30\%$ of the flux of \rxj, with the compact nebula plus jet comprising the remaining  $\sim70\%$. The luminosity of the fainter large-scale emission
 (within $r<4'$ from the pulsar)   is about a factor of  5--10  larger than that of the compact PWN and pulsar.  
The X-ray spectrum of the point source can be described by an absorbed power-law
 plus blackbody model, with a photon index of $\Gamma = 1.6 \pm 0.6$.  The compact PWN  spectrum 
 is harder, with a photon index 
 $\Gamma\simeq1-1.3$  \citep{Halpern2004}. 
The spectrum of the large-scale diffuse emission was fitted by \citet{Caraveo2010} with a power-law modified by the interstellar absorption. The fit gave $\Gamma = 1.8 \pm 0.1$.  However, the fit quality was fairly poor suggesting a more complex spectrum (e.g., a possible additional thermal component; see below).

\section{Discussion}\label{sec:discussion}

We have discovered spatially extended TeV emission from the region of CTA~1. Here we 
 discuss the results in the context of the available multiwavelength data. 

\pagebreak

\subsection{The nature of the TeV source: A PWN scenario}
 
The good positional match between \ver\ and \psr\ within the CTA~1 SNR makes their physical association virtually indisputable, while leaving uncertain which component powers the TeV emission.  The source is unlikely to be  related to a gamma-ray binary or a background blazar given that the TeV emission is extended and non-variable. 
 However, the extent of \ver\ is also much smaller than that of the SNR and, hence,  the TeV source does not resemble several  SNRs whose shells have been resolved in TeV gamma rays \citep{Komin2011}. 
 There still remains a possibility that only part of the CTA~1 SNR shell is interacting with a dense molecular cloud which could cause a local enhancement in the TeV brightness of the emission~\citep{Komin2011}.  
 However, we do not find any evidence for such a cloud at any other wavelength, including 60~$\mu$m IR  or HI  \citep[see][]{Pineault1993},  and the high Galactic latitude of the SNR places it nearly 250~pc above the Galactic plane, much higher than the scale height of molecular clouds. 
 
\ver\ must then be 
powered by the young \psr.  Extrapolating the pulsed emission seen by the \fermi-LAT \citep{Abdo2012CTA1}, assuming a power-law above the break energy of $\sim4$~GeV, gives an estimated TeV flux 2--3 orders of magnitude below the observed flux. 
Therefore, the most plausible remaining scenario is that the TeV emission originates from the X-ray PWN surrounding the pulsar. The PWN 
 consists of a bent jet, a compact core, and a large-scale diffuse component, as seen in the \chandra\ and \xmm\  
images (see Figure~\ref{fig:xray_pwn}).
 The bending 
 of the jet (see the inset in Figure~\ref{fig:xray_pwn}, and also \citealt{Halpern2004}) 
 could be caused by the ram pressure of the oncoming medium due to the 
 NW--SE pulsar motion or by the interaction with a reverse shock propagating 
 NW within the SNR extent.
 Alternatively, a kink instability might be responsible for the bending of the jet. Indeed, the Vela pulsar jet shows some kink-like shape changes. However, these wiggles tend to occur on smaller spatial scales, while globally the Vela jet is always (during the last 10 years) seen to bend toward one side, likely due to the pressure of the oncoming ambient material \citep{Pavlov2003}.
 Similarly, in the \xmm\ and \chandra\ images of CTA~1, obtained at different epochs, the jet is seen to consistently bend in the same direction. 
  
 If the change in the jet morphology is caused by ram pressure, then we can estimate the pressure from the jet's curvature, following  \citet{Pavlov2003}. Assuming that the jet pressure is dominated by the contribution from a magnetic field $B_{-4} = B/(10^{-4}$~G),  for a jet curvature radius $R_{\rm curv}\simeq10''$ and a jet diameter $d_{\rm jet}\sim1''$ the pressure estimate is $P_{\rm ram}\sim1\times 10^{-11}B_{-4}^2$  erg cm$^{-3}$.
    Despite the fairly high magnetic field assumed, the estimated pressure is rather  low compared to the ambient pressure inferred  for other young 
   pulsars with X-ray PWNe resolved by \chandra\ \citep{Kargaltsev2008}. Assuming that  the ram pressure is caused by the pulsar motion through a medium of density $n_{-1} = n/(10^{-1}~{\rm cm}^{-3})$, one obtains 
a very modest  pulsar speed 
$v\simeq90\ n_{-1}^{-1/2}B_{-4}$ km~s$^{-1}$ which corresponds to the proper motion of just 
$\sim0.013\ n_{-1}^{-1/2}B_{-4} d_{1.4}^{-1}$ arcs yr$^{-1}$, 
 assuming a distance of $d_{1.4}= d/(1.4$~kpc). 
    This means that over its lifetime $\tau$ the pulsar should have moved by  only 
  $\sim3\ n_{-1}^{-1/2}  (\tau/1.3\times 10^{4}~{\rm yrs}) B_{-4} d_{1.4}^{-1}$
   arcminutes.   This distance is much less than the size of the SNR and the extent of the TeV source, and it is  even less than the extent of large-scale X-ray PWN seen in the \xmm\ and \asca\ images. 
   We also note that despite being dependent on several parameters, the  above estimate of the distance traveled by the pulsar likely 
 represents  an upper limit. 
Thus the estimated velocity from the jet-bending is inconsistent with the otherwise plausible
hypothesis that the NW 
extension of the TeV source (see Figure~\ref{fig:hardextended_radioexcess})
might be due to 
aged relativistic electrons left behind by the fast moving pulsar.
   
  It is also possible that the relic PWN has been pushed to one  side (i.e.\ NW of the pulsar) by the reverse shock that must have arrived from the SE direction.  Indeed,  such a scenario is  supported by the overall asymmetry of the SNR shell 
   which appears to expand into much lower density medium in its NW part and hence the reverse shock is not expected to arrive from the NW direction. 
A recent 
interaction with the reverse shock could possibly also explain the bending of the jet while emission ahead of the pulsar 
could be explained by the turbulent mixing between the pulsar wind and SN ejecta behind the reverse shock (similar to G327.1-1.1; \citealt{Temim2009}).
The latter may contribute some thermal emission and explain the poor quality of the power-law fit to the extended X-ray emission \citep{Caraveo2010}.
A deep mapping of the entire SNR by \xmm\ 
can  test this hypothesis by providing a high-S/N spectrum of the faint large-scale X-ray emission which should then contain a thermal emission component coming from the ejecta (c.f.\ e.g., Vela X spectrum; \citealt{LaMassa2008}). 

    Pulsar wind particles may be transported either by diffusion or by advection. One could in principle compare the two terms, if the bulk flow speed (as a function of distance from the pulsar) and the magnetic field structure were known. The MHD models for isotropic pulsar winds \citep[e.g.,][]{Kennel84}
are unlikely to be valid on large scales and when 
mixing due to interaction with 
the reverse shock is present. 
However, we can make some estimates of the average magnetic field by assuming which process is dominant in transporting the particles.
   
   Assuming that X-ray- and TeV-emitting particles move away from the pulsar with similar velocities (i.e.\ that the X-ray- and TeV-emitting regions are co-spatial and that the effects of energy-dependent diffusion are negligible),  
and that the synchrotron cooling-time is the dominant time-scale, one can  crudely estimate the magnetic field strength \citep[see, e.g.,][]{Aharonian2005b}. For X-ray and TeV gamma-ray emission regions of sizes $R_{\rm X}$ and $R_\gamma$, respectively, with $\overline{E_{\rm X}}$ and  $\overline{E_{\gamma}}$ being the corresponding mean energies of the 
photons in keV and TeV units, the magnetic field is $B_{\rm pwn}\sim160(R_{\rm X}/R_{\gamma})^2(\overline{E_{\rm X}}/\overline{E_{\gamma}})$~$\mu$G.
 For the observed  $R_{\rm X}/R_{\gamma}\approx0.5$, 
 $ \overline{E_{\rm X}}=5$~keV, and $\overline{E_{\gamma}}=5$~TeV, the corresponding average magnetic field is $\sim40$~$\mu$G.
This is much higher than what is suggested by modeling (see below) and also much higher than what is seen in other such evolved systems.

Note that the ratio $R_{\rm X}/R_{\gamma}\approx0.5$ is likely an underestimate and the \asca\ data suggest that it can be a factor of 2--3 
larger 
 (see Fig.~\ref{fig:ASCA_xray}).  Indeed, in the \asca\ images some diffuse emission appears to be seen up to $40'$ away from the pulsar \citep{Slane2004}.
 The true extent and the nonthermal nature of the faint X-ray emission can only be measured in deep observations with \xmm.
Similarly, the TeV size we quote is a lower limit since more the extended portions away from the pulsar are likely to be the fainter than our detection threshold. 
The estimate of the average magnetic field should thus be taken 
 as a crude order-of-magnitude estimate.

 Another estimate of the magnetic field can be made by assuming that diffusion is the dominant transport mechanism throughout the nebula.
In the simplest case of cross field diffusion (the Bohm limit), the diffusion constant is given by $D=\gamma mc^3/3 eB$, where $\gamma$ is the electron Lorentz factor. 
Using the relation $E_\gamma\sim \gamma^2\epsilon$,  
where $E_\gamma$ is the mean up-scattered IC photon energy and $\epsilon$ is the seed photon energy, the diffusion constant for electrons scattering on the CMB in a magnetic field $B_{-5} = B_{\rm pwn}/(10^{-5}$~G) can then be expressed as $D = 8.5\times10^{25}E_{\gamma}^{1/2}B_{-5}^{-1}  {\rm\ cm}^{2} {\rm\ s}^{-1}$.
  Assuming that particles  travel during their characteristic cooling time of $\tau_{\gamma}\approx100 (1+14.4B_{-5}^2)^{-1}E_{\gamma}^{-1/2}$~kyrs \citep[see][]{deJager2009},
   the diffusion length is  $\sim(6 D \tau_{\gamma})^{1/2}~\sim 13 B_{-5}^{-1/2} (1+14.4B_{-5}^2)^{-1/2}$~pc which translates into  $\sim20' d_{1.4}^{-1}$ for $B_{\rm pwn}=5$~$\mu$G.  (Note that, for a given distance to CTA~1, this estimate depends only on the magnetic field strength.) 
This size  roughly corresponds to the observed extent of the TeV source.
%
Although small, such low magnetic field ($\sim5$~$\mu$G)  was inferred  through the multiwavelength spectral modeling for the Vela~X  plerion \citep{deJager2008}. 
The  low $B_{\rm pwn}$ resulting from the Bohm diffusion estimates has been previously noticed for several other relic PWNe (e.g.\ \citet{deJager2009}, as well as by \citet{Uchiyama2009} and \citet{Anada2010} based on their analysis  of the synchrotron spectra measured by {\em Suzaku} across the extent of TeV sources  HESS~J1825--137  and HESS~J1809--193.)

\begin{figure}[tbp]
\begin{center}
\includegraphics[width=\columnwidth,angle=0]{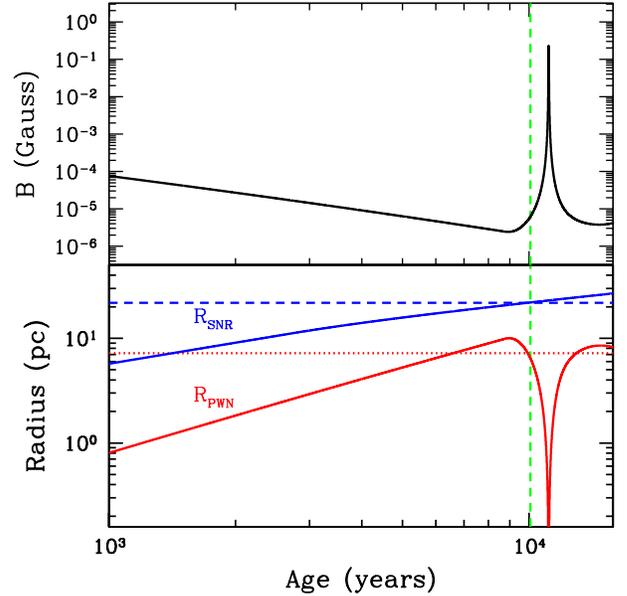} 
\end{center}
\vspace{-0.4cm}
\caption{Upper: Time evolution of the 
PWN magnetic field, using the model of \citet{Gelfand2009}, with parameters given in Table~\ref{tab:cta1_modelParams}. Lower: Time evolution of the modeled SNR (blue) and PWN (red) radii. Horizontal dashed lines indicate the current values for CTA 1. The vertical 
 green line indicates the age at which the measured SNR radius is reached. (See text for model description.)}
\label{fig:CTA1SlaneRadii} 
\end{figure}

Furthermore, we can use a 
 dynamical model for the evolution of a PWN inside a non-radiative SNR \citep{Gelfand2009}  to estimate the physical properties of the PWN. 
The results are illustrated in Figure~\ref{fig:CTA1SlaneRadii}.  
We find that to correctly reproduce the radius of the SNR (lower panel, shown in blue) while simultaneously matching the estimated PWN radius (shown in red), 
 the current spin-down power, and the total TeV flux, we require an ambient density $n_0 \approx 0.07 {\rm\ cm}^{-3}$. 
 This is somewhat 
  larger than that estimated from \asca\ measurements of the thermal X-ray emission \citep{Slane1997, Slane2004}, although those measurements were based on observations of a small fraction of the SNR shell.

\begin{figure}[tbp]
\begin{center}
\includegraphics[width=\columnwidth,angle=0]{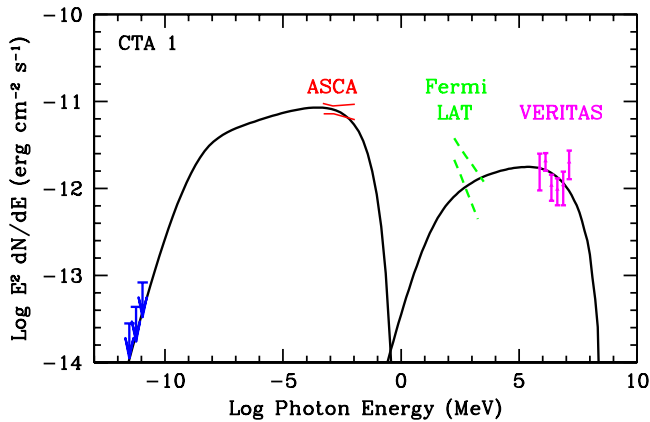} 
\end{center}
\vspace{-0.4cm}
\caption{Broadband emission from CTA~1, along with the synchrotron and inverse Compton emission from the PWN predicted by the model of \citet{Gelfand2009} with parameters given in Table~\ref{tab:cta1_modelParams}. See the text for the derivation of the 1.4~GHz radio upper limits shown in blue.  The red lines mark the errors on the X-ray spectrum measured by \asca\ \citep{Slane2004}.  The green, dashed lines mark the errors on the unpulsed  \fermi-LAT spectrum \citep{Abdo2012CTA1}.}
\label{fig:CTA1SlaneSED} 
\end{figure}

Also shown in Figure~\ref{fig:CTA1SlaneRadii} (upper panel) is the time evolution of the PWN magnetic field for this model. At the current age of $\sim 10$~kyr 
 implied by the SNR radius, the PWN magnetic field strength is $\sim 6 \mu$G. The recent decrease in the PWN radius, and increase in the magnetic field, result from the beginning of the SNR reverse shock interaction with the nebula, as suggested by other arguments presented above.
Figure~\ref{fig:CTA1SlaneSED} shows the archival broadband data for CTA~1 along with the emission predicted for the model used in Figure~\ref{fig:CTA1SlaneRadii} assuming a broken
 power law injection of particles from the pulsar, for which a braking index of 3 is assumed. The model parameters are summarized in Table~\ref{tab:cta1_modelParams}.

Radio observations of CTA~1 do not provide conclusive evidence for emission from the PWN \citep{Pineault1997}. We note that, in their
modeling, \citet{Zhang2009} assumed that the entire emission from the SNR was associated with the PWN. In fact, the PWN is much
fainter.  Here we have used the 1.4~GHz image from \citet{Pineault1997} to estimate the flux within a 20~arcminute radius around the
pulsar, and have used this flux as an upper limit for the PWN emission. In Figure~\ref{fig:CTA1SlaneSED}, we have extrapolated this upper limit to
lower frequencies assuming a spectral index $\alpha = 0.3$ (where $S_\nu \propto \nu^{-\alpha}$, is the flux at the frequency $\nu$) and to higher frequencies assuming
$\alpha = 0.$ These index values represent the typical range observed in radio PWN, and the associated flux values correspond to conservative
upper limits. 

The TeV spectrum predicted by \citet{Zhang2009} overpredicts the TeV
flux and requires an initial spin period in the range $30-40$~ms. 
In comparison, the model shown in Figure~\ref{fig:CTA1SlaneSED} agrees with the
\VERITAS\ data, within uncertainties, and we conclude $P_0 \sim 155$~ms. This difference could arise
from the fact that \citet{Zhang2009} assign the full radio flux from CTA~1 to the PWN, thus requiring significantly more energy in the electron spectrum. 
An additional difference in the models is that we have calculated the evolving magnetic field strength based on the
fraction of spin-down energy injected as magnetic flux whereas \citet{Zhang2009} assume a time-dependent field value which 
is independent
of any other system parameters. Our results suggest a break energy of $\sim 50$~GeV 
with $\sim 80\%$ of the spin-down power appearing
in the form of particle flux. Like most such systems, CTA~1 is thus a particle-dominated PWN.

The dashed green curves in Figure~\ref{fig:CTA1SlaneSED} represent the best fit for the unpulsed \fermi-LAT spectrum published in \citet{Abdo2012CTA1}. In the model calculation shown, the TeV emission is produced by inverse Compton scattering of photons from the CMB, integrated starlight, and infrared emission from local dust, following the approximate prescription given by \citet{Strong2000}. 
 The model produces reasonable agreement with the radio, X-ray, and TeV data with a solution that gives approximately the correct pulsar spin-down power and characteristic age at the current epoch. However, the model is a poor fit to the reported \fermi-LAT spectrum. We have considered additional photon fields to produce enhanced inverse Compton emission at GeV energies, but have been unable to reproduce the published spectral index.  We note that the reported unpulsed GeV emission is quite faint and it is in the presence of bright pulsed emission from \psr. It will be of considerable interest to carry out further investigations of this unpulsed emission as more \fermi-LAT data are accumulated.

\begin{table}[tbp]
\caption{Model parameters for broadband emission from CTA~1. See \citet{Gelfand2009} for parameter definitions.}
\vspace{-0.3cm}
\begin{center}
\begin{tabular*}{\columnwidth}{@{\extracolsep{\fill}}lc}
\hline
Parameter & Value \\ \hline
\multicolumn{2}{c}{Input:} \\
Explosion Energy, $E_{SN}$	&	$10^{51}$ erg (fixed)	\\
Ejecta Mass, $M_{ej}$		&	$6.1 {\rm\ M_\odot}$		\\
Ambient density, $n_0$		&	$0.068 {\rm\ cm}^{-3}$ 	\\
Initial spin-down, $\dot{E_0}$	&	$7.5 \times 10^{36}{\rm\ erg\ s}^{-1}$	 \\
Spin-down timescale, $\tau_0$	&	$3.2 \times 10^{3}$~yr		\\
Braking index, $n$			&	3 (fixed)	\\
$\eta_B$					&	0.2	\\
$\alpha_1$				&	0.5		\\
$\alpha_2$				&	2.8		\\
Break energy, $E_b$		&	50 GeV	\\
 \hline
 \multicolumn{2}{c}{Output:} \\
Age					&	$1.0 \times 10^{4}$~yr 	\\
$B_{PWN}$			&	$6.3 {\rm\ \mu G}$ 		\\
$\dot{E}$				&	$4.4 \times 10^{35}{\rm\ erg\ s}^{-1}$ \\
$\tau_c$				&	$1.3 \times 10^{4}$~yr	\\
$P_0$				&	155 ms	\\
 \hline
\end{tabular*}
\end{center}
\label{tab:cta1_modelParams} 
\end{table}


\subsection{Comparison with other relic PWNe}

\begin{figure}[tbp]
\begin{center}
\includegraphics[angle=90,width=\columnwidth]{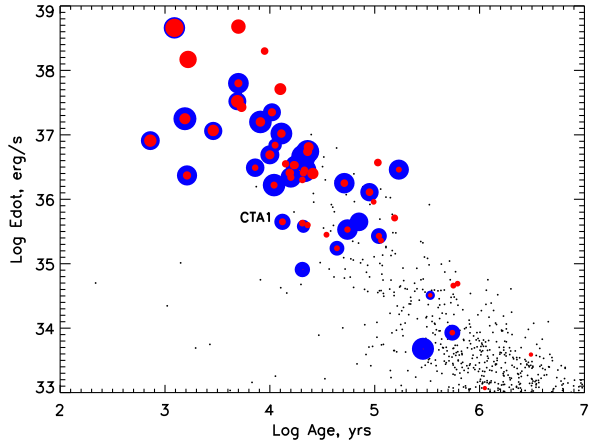}
\caption{Pulsar spin-down luminosity vs age, from \citet{Kargaltsev2010}, with CTA~1 point indicated. 
Filled circles: X-ray (red) and TeV (blue) luminosities of PWNe or PWN candidates. Larger circle sizes correspond to higher luminosities in the corresponding waveband. Small black dots denote ATNF catalog pulsars \citep{ATNF}\protect\footnotemark.}
\label{fig:CTA1_EdotVsAge}
\end{center}
\end{figure}

\begin{figure}[htbp]
\begin{center}
\hspace{-1cm}
\includegraphics[angle=90,width=\columnwidth]{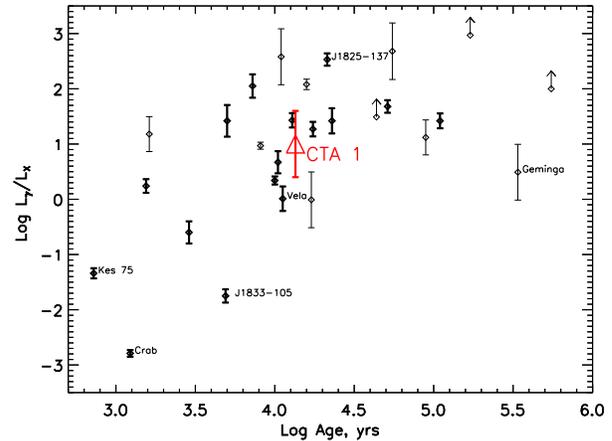}
\caption{Plot of the ratio of TeV to X-ray luminosity vs pulsar spin-down age, from \citet{Kargaltsev2010}, with CTA~1 shown by the red triangle. Thick and thin error bars correspond to firm and tentative (or questionable) PWN associations, with selected objects labeled (see \citeauthor{Kargaltsev2010} for further details).}
\label{fig:CTA1_LgVsAge}
\end{center}
\end{figure}

Figures~\ref{fig:CTA1_EdotVsAge} and~\ref{fig:CTA1_LgVsAge} 
show the comparisons of the properties of CTA~1 with other PWNe and PWNe candidates detected at 
TeV energies. At the  distance of 1.4~kpc, the $>1$~TeV luminosity of the PWN in CTA~1 is $9.4\times10^{32}$ erg s$^{-1}$.  
Figure~\ref{fig:CTA1_EdotVsAge} shows the relative luminosities of PWNe in the TeV and X-ray bands, as functions of spin-down power and characteristic age \citep{Kargaltsev2010}. CTA~1 fits 
with the picture that TeV PWNe are generally found around pulsars with ages $\lesssim$100~kyrs and $\dot{E}\gtrsim10^{35}$~erg s$^{-1}$, 
 although the TeV luminosities do not depend on the pulsar age nearly as much  as the X-ray PWNe luminosities do. 
Figure~\ref{fig:CTA1_LgVsAge} shows the distance-independent ratio of the TeV to X-ray luminosity as a function of characteristic age for a set of PWNe or PWN candidates, with 
CTA~1 marked by the red triangle.  
Additionally, the estimated size of the TeV emission region in CTA~1 is consistent with the sizes of TeV nebulae around pulsars with ages similar to that of \psr, although the large errors on the estimated distances prohibit a definite correlation. 
A comparison of CTA~1 with the TeV/X-ray PWN population supports the  PWN origin of the TeV emission.

\section{Summary and Conclusions}
\veritas\ has detected TeV gamma-ray emission coincident with \snr. The emission is extended, with a centroid near the \fermi\ gamma-ray pulsar \psr\ and its X-ray pulsar wind nebula (PWN). The photon spectrum is well described by a power-law with differential spectral index of $\Gamma= 2.2\pm 0.2_{\rm stat} \pm 0.3_{\rm sys}$ and an integral flux above 1~TeV corresponding to $\sim 4\%$ of the steady TeV gamma-ray emission from the Crab Nebula. 
It is unlikely that the TeV emission is due to interaction of the CTA~1 shell with a dense molecular cloud, given the 
lack of evidence for such a cloud at other wavelengths (60~$\mu$m IR or HI maps). We have analyzed archival X-ray data from \rosat\ (0.5--2~keV) and \asca\ (4--10~keV) of the large scale nebula and \xmm\ (0.5--10~keV) and \chandra\ (0.5--8~keV) of the region close to the pulsar 
and find that the large scale emission seems to match the TeV morphology. 
The positional coincidence with the pulsar, small extent of the TeV emission, and X-ray/TeV luminosities strongly argue for a PWN origin. 
We have estimated the magnetic field strength 
assuming particle transport by either diffusion or by advection. A more detailed dynamical model of the SNR-PWN system suggests a 6~$\mu$G field along with a recent interaction between the PWN and the SNR reverse shock.

\footnotetext{http://www.atnf.csiro.au/research/pulsar/psrcat}

\par\vspace{\baselineskip}

This research is supported by grants from the U.S. Department of Energy Office of Science, the U.S. National Science Foundation and the Smithsonian Institution, by NSERC in Canada, by Science Foundation Ireland (SFI 10/RFP/AST2748) and by STFC in the U.K. We acknowledge the excellent work of the technical support staff at the Fred Lawrence Whipple Observatory and at the collaborating institutions in the construction and operation of the instrument.

O.K. was supported through the National Science Foundation grant No. AST0908733.  P.S. acknowledges support from NASA contract NAS8-03060.

\begin{thebibliography}{99}

\bibitem[Abdo \etal(2008)]{Abdo2008CTA1}Abdo, A.~A., \etal\ 2008, Science, 322, 1218

\bibitem[Abdo \etal(2010)]{Abdo2010Geminga}Abdo, A.~A., \etal\ 2010, ApJ, 712, 1209

\bibitem[Abdo \etal(2012)]{Abdo2012CTA1}Abdo, A.~A., \etal\ 2012, ApJ, 744, 146

\bibitem[Acciari \etal(2008)]{Acciari2008}Acciari, V.~A., \etal\ 2008, ApJ, 679, 1427

\bibitem[Aharonian \etal(1997)]{Aharonian1997}Aharonian, F.~A., \etal\ 1997, Astroparticle Physics, 6, 343

\bibitem[Aharonian \etal(2005a)]{Aharonian2005}Aharonian, F., \etal\ 2005a, Science, 307, 1938

\bibitem[Aharonian \etal(2005b)]{Aharonian2005b} Aharonian, F.~A.,  \etal\ 2005b, A\&A, 442, L25 

\bibitem[Aharonian \etal(2006a)]{Aharonian2006}Aharonian, F., \etal\ 2006a, ApJ, 636, 777, arXiv:astro-ph/0510397v1

\bibitem[Aharonian \etal(2006b)]{Aharonian2006b}Aharonian, F., \etal\ 2006b, A\&A, 457, 899

\bibitem[Anada \etal(2010)]{Anada2010} Anada, T., \etal\ 2010, PASJ, 62, 179

\bibitem[Atoyan \& Aharonian (1996)]{Atoyan1996}Atoyan, A.~M. \& Aharonian, F.~A. 1996, MNRAS, 278

\bibitem[Berge, Funk \& Hinton(2007)]{Berge2007}Berge, D., Funk, S., \& Hinton, J. 2007, A\&A, 466, 1219

\bibitem[Brazier \etal(1998)]{Brazier1998}Brazier, K.~T.~S., \etal\ 1998, MNRAS, 295, 819

\bibitem[Caraveo \etal(2010)]{Caraveo2010}Caraveo, P.~A., \etal\ 2010, ApJ Letters, 725, 6

\bibitem[Cogan (2008)]{Cogan2008}Cogan, P. 2008, in Proc 30th ICRC, Merida, Mexico, 3, 1385

\bibitem[Daniel (2008)]{Daniel2008}Daniel, M.~K. 2008, in Proc 30th ICRC, Merida, Mexico, 3, 1325

\bibitem[de Jager, Slane \& LaMassa(2008)]{deJager2008}de~Jager, O., Slane, P.~O., LaMassa, S. 2008, ApJ, 689, L125 


\bibitem[de Jager \& Djannati-Ata{\"i}(2009)]{deJager2009} de~Jager, O.~C., \& Djannati-Ata{\"i}, A.\ 2009, Astrophysics and Space Science Library, 357, 451 

\bibitem[Fomin \etal(1994)]{Fomin1994}Fomin, V.~P., \etal\ 1994, Astroparticle Physics, 2, 137

\bibitem[Gaensler \& Slane (2006)]{Gaensler2006}Gaensler B.~M. \& Slane P.~O. 2006, ARAA, 44, 17
 
 \bibitem[Gelfand, Slane \& Zhang (2009)]{Gelfand2009}	Gelfand, J.~D, Slane, P.~O. \& Zhang, W. 2009, ApJ, 703, 2051. 

\bibitem[Hall \etal(2001)]{Hall2001}Hall, T.~A., \etal\ 2001, in Proc 27th ICRC, Hamburg, Germany

\bibitem[Halpern \& Holt (1992)]{Halpern1992}Halpern, J.~P. \& Holt, S.~S. 1992, Nature, 357, 222

\bibitem[Halpern, Camilo, \& Gotthelf (2007)]{Halpern2007}Halpern, J.~P., Camilo, C., Gotthelf, E.~V. 2007, ApJ, 668, 1154  

\bibitem[Halpern \etal(2004)]{Halpern2004}Halpern, J.~P., \etal\ 2004, ApJ, 612, 398 

\bibitem[Harris \& Roberts (1960)]{HarrisPASP1960}Harris, D.~E., \& Roberts, J.~A. 1960, PASP, 72, 347. 

\bibitem[Hartman \etal(1999)]{Hartman1999}Hartman, R.~C., \etal\ 1999, ApJS, 123, 79

\bibitem[Hillas (1985)]{Hillas1985}Hillas, M. 1985, in Proc 19th ICRC, La Jolla, IL, 3, 445

\bibitem[Hinton \& Hofmann(2010)]{Hinton2010}Hinton, J.~A. \& Hofmann, W. 2010, AARA, 47, 523


\bibitem[Holder \etal(2011)]{Holder2011}Holder, J., \etal\ 2011, in Proc 32nd ICRC, Beijing, arXiv:1111.1225v1 

\bibitem[Kargaltsev \& Pavlov(2008)]{Kargaltsev2008}Kargaltsev, O. \&  Pavlov,  G.~G. 2008, AIP Conf. Proc., 983, 171

\bibitem[Kargaltsev \& Pavlov(2010)]{Kargaltsev2010}Kargaltsev, O. \& Pavlov, G.~G. 2010, AIP Conf. Proc., 1248, 25

\bibitem[Kennel  \& Coroniti(1984)]{Kennel84} Kennel, C.~F., \& Coroniti, F.~V.\ 1984, ApJ, 283, 694 

\bibitem[Khelifi \etal(2001)]{Khelifi2001}Khelifi, B., \etal\ 2001, in Proc 27th ICRC, Hamburg, Germany

\bibitem[Komin \etal(2011)]{Komin2011}Komin, N., \etal 2011, in Proc 23rd Rencontres de Blois

\bibitem[Krawczynski \etal(2006)]{Krawczynski2006}Krawczynski, H., \etal\ 2006, Astroparticle Physics, 25, 380 

\bibitem[LaMassa, Slane \& de Jager(2008)]{LaMassa2008} LaMassa, S.~M., Slane, P.~O., \& de Jager, O.~C.\ 2008, ApJ, 689, L121

\bibitem[Lemiere \etal(2009)]{Lemiere2009}Lemiere, A., \etal\ 2009, ApJ, 706, 1269

\bibitem[Li \& Ma(1983)]{LiMa1983}Li, T.-P. \& Ma, Y.-Q. 1983, ApJ, 272, 317

\bibitem[Manchester \etal(2005)]{ATNF}Manchester, R.~N., \etal\ 2005, Astronomical Journal, 129, 1993

\bibitem[Mattana \etal(2009)]{Mattana2009} Mattana, F., \etal\ 2009, ApJ, 694, 12

\bibitem[Montmerle (1979)]{Montmerle1979} Montmerle, T. 1979, ApJ, 231, 95 

\bibitem[Ong \etal(2009)]{Ong2009}Ong, R.~A., \etal\ 2009,  in Proc 31st ICRC, Lodz, Poland, arXiv:0912.5355

\bibitem[Pavlov \etal(2003)]{Pavlov2003}Pavlov, G.~G., \etal\ 2003, ApJ, 591, 1157


\bibitem[Pineault \etal(1993)]{Pineault1993} Pineault, S., \etal\ 1993, AJ, 105, 1060

\bibitem[Pineault \etal(1997)]{Pineault1997}Pineault, S., \etal\ 1997, A\&A, 324, 1152

\bibitem[Roberts, Romani \& Kawai(2001)]{Roberts2001}Roberts, M.~S.~E., Romani, R.~W., \& Kawai, N. 2001, ApJ, 133, 451

\bibitem[Rowell \etal(2003)]{Rowell2003}Rowell, G., \etal\ 2003,   in Proc. 28th ICRC, Tokyo, Japan

\bibitem[Seward, Schmidt \& Slane(1995)]{Seward1995}Seward, F.~D., Schmidt, B., Slane, P. 1995, ApJ, 453, 284

\bibitem[Sieber, Salter, \& Mayer (1981)]{Sieber1981}Sieber, W., Salter, C.~J., \& Mayer, C.~J. 1981, A\&A, 103, 393

\bibitem[Slane \etal(1997)]{Slane1997}Slane, P., \etal\ 1997, ApJ, 485, 221

\bibitem[Slane \etal(2004)]{Slane2004}Slane, P., \etal\ 2004, ApJ, 601, 1045

\bibitem[Strong, Moskalenko, \& Reimer(2000)]{Strong2000}Strong, A.~W., Moskalenko, I.~V., \& Reimer, O. 2000, ApJ, 537, 763

\bibitem[Temim \etal(2009)]{Temim2009} Temim, T., \etal\ 2009, ApJ, 691, 895 

\bibitem[Torres \etal(2003)]{Torres2003}Torres, D.~F., \etal\ 2003, Physics Reports 382, 303

\bibitem[Uchiyama \etal(2009)]{Uchiyama2009}Uchiyama, H., \etal\ 2009, PASJ, 61, S189

\bibitem[Weekes \etal(1989)]{Weekes1989}Weekes, T.~C., \etal\ 1989, ApJ, 342, 379


\bibitem[Zhang, Jiang \& Lin (2009)]{Zhang2009}Zhang, L., Jiang, Z.~J., \& Jin, G.~F. 2009, ApJ, 699, 507

\end{thebibliography}
\end{document}